\documentclass[12pt]{iopart}
\usepackage{iopams}  
\usepackage{graphicx, caption}
\usepackage{color}
\usepackage{cite}
\usepackage{comment}
\usepackage{hyperref}
\usepackage{ulem}
\definecolor{orange}{rgb}{0.9,0.5,0}
\definecolor{dodgerblue}{rgb}{0.12, 0.56, 1.0}

\newcommand{\newtxt}[1]{#1}

\usepackage{acronym}

\newacro{adm}[ADM]{Arnowitt-Deser-Misner}
\newacro{amr}[AMR]{adaptive mesh refinement}
\newacro{bbh}[BBH]{binary black hole}
\newacro{bh}[BH]{black hole}
\newacroplural{bh}[BHs]{black holes}
\newacro{bhns}[BHNS]{black hole-neutron star}
\newacro{bhpt}[BHPT]{black hole perturbation theory}
\newacro{bns}[BNS]{binary neutron star}
\newacro{bf}[BF]{Bayes' factor}
\newacro{cbc}[CBC]{compact binary coalescence}
\newacro{cce}[CCE]{Cauchy characteristic extraction}
\newacro{ce}[CE]{Cosmic Explorer}
\newacro{da}[DA]{data analysis}
\newacro{et}[ET]{Einstein Telescope}
\newacro{eob}[EOB]{Effective-One-Body}
\newacro{eom}[EOM]{equations of motion}
\newacro{fd}[FD]{frequency domain}
\newacro{fft}[FFT]{Fast Fourier transform}
\newacro{gw}[GW]{gravitational-wave}
\newacroplural{gw}[GWs]{Gravitational-waves}
\newacro{gr}[GR]{general relativity}
\newacro{grb}[GRB]{gamma-ray burst}
\newacro{grhd}[GRHD]{general-relativistic hydrodynamics}
\newacro{gwosc}[GWOSC]{Gravitational Wave Open Science Center}
\newacro{gwtc1}[GWTC-1]{the first gravitational-wave transients catalog}
\newacro{gsf}[GSF]{Gravitational Self Force}
\newacro{hm}[HM]{Higher mode}
\newacroplural{hm}[HMs]{Higher modes}
\newacro{ifo}[IFO]{interferometer}
\newacro{imr}[IMR]{inspiral-merger-ringdown}
\newacro{im}[IMR]{inspiral-to-merger}
\newacro{kagra}[KAGRA]{Kamioka Gravitational Wave Detector}
\newacro{ligo}[LIGO]{Laser Interferometer Gravitational-Wave Observatory}
\newacro{lisa}[LISA]{Laser Interferometer Space Antenna}
\newacro{lr}[LR]{Light Ring}
\newacro{lso}[LSO]{Last Stable Orbit}
\newacro{lvc}[LVC]{LIGO-Virgo Collaboration}
\newacro{lvk}[LVK]{LIGO-Virgo-Kagra Collaboration}
\newacro{lo}[LO]{leading order}
\newacro{ns}[NS]{neutron star}
\newacroplural{ns}[NSs]{neutron stars}
\newacro{nr}[NR]{numerical relativity}
\newacro{nqc}[NQCs]{Next-to-quasicircular corrections}
\newacro{nlo}[NLO]{next-to-leading order}
\newacro{nnlo}[NNLO]{next-to-next-to-leading order}
\newacro{n3lo}[N3LO]{next-to-next-to-next-to-leading order}
\newacro{n4lo}[N3LO]{next-to-next-to-next-to-next-to-leading order}
\newacro{ode}[ODE]{Ordinary Differential Equation}
\newacroplural{ode}[ODEs]{Ordinary Differential Equations}
\newacro{pn}[PN]{post-Newtonian}
\newacro{pm}[PM]{post-Minkowskian}
\newacro{pe}[PE]{parameter estimation}
\newacro{psd}[PSD]{power spectral density}
\newacro{qnm}[QNM]{quasi-normal mode}
\newacro{qc}[QC]{quasi-circular}
\newacro{snr}[SNR]{signal-to-noise ratio}
\newacro{sxs}[SXS]{Simulating eXtreme Spacetimes}
\newacro{td}[TD]{time domain}
\newacro{ng}[NG]{Nect Generation}

\begin{document}
\title{AthenaK simulations of the binary black hole merger GW150914}

\author{David Radice$^{1,2,3}$\footnote{Alfred P.~Sloan Fellow}, 
Rossella Gamba$^{1,2,4}$,
Hengrui Zhu$^{5,6}$,
and Alireza Rashti$^{1,2}$
}

\address{${}^1$ Institute for Gravitation \& the Cosmos, The Pennsylvania State University, University Park, PA 16802}
\address{${}^2$ Department of Physics, The Pennsylvania State University, University Park, PA 16802}
\address{${}^3$ Department of Astronomy \& Astrophysics, The Pennsylvania State University, University Park, PA 16802}
\address{${}^4$ Department of Physics, University of California, Berkeley, CA 94720}
\address{${}^5$ Department of Physics, Princeton University, Jadwin Hall, Washington Road, New Jersey, 08544, USA}
\address{${}^6$ Princeton Gravity Initiative, Princeton University, Princeton, New Jersey, 08544, USA}

\ead{dur566@psu.edu}

\begin{abstract}
We present new binary black hole simulations targeted to GW150914 using the GPU-accelerated code \texttt{AthenaK}. We compute the properties of the final remnant with the isolated horizon formalism and obtain gravitational-waveforms at future null infinity via Cauchy characteristic extraction. We compare our results with those obtained by the Simulating eXtreme Spacetimes (SXS) and Rochester Institute of Technology (RIT) groups, targeted to the same event. We find excellent agreement with the SXS and RIT results in the remnant mass, spin, and recoil velocity. For the dominant $(\ell,m)=(2,2)$ mode of the gravitational-wave signal we find maximum dephasing of $\Delta \phi \simeq 0.35$ and amplitude difference of $\Delta A/A \simeq 0.4\%$. We use our newly computed waveform to re-analyze the GW150914 data and find posteriors for chirp mass, luminosity distance, and inclination that are broadely consistent with those obtained using semi-analytic waveform models. This work demonstrates the viability of \texttt{AthenaK} for many-orbits binary black hole merger simulations. A step-by-step tutorial, including all necessary input files and analysis scripts to reproduce our results, is available on \texttt{GitHub}.
\end{abstract}

\newpage
\section{Introduction}
\label{sec:introduction}
The detection of the \ac{gw} signal GW150914 from a \ac{bbh}\acused{bh} merger by the LIGO observatories in Hanford (WA) and Livingston (LA) opened a new window on the Universe and ushered in the era of \ac{gw} astronomy \cite{LIGOScientific:2016aoc, LIGOScientific:2016vlm, LIGOScientific:2016vpg}. This discovery was recognized with the 2017 Physics Nobel Prize to Rainer Weiss, Barry C.~Barish, and Kip S. Thorne. GW150914 confirmed the predictions of Einstein's general relativity in the strong field regime \cite{LIGOScientific:2016lio, Cardoso:2016rao, Yunes:2016jcc, Cardoso:2019rvt}, for example by enabling tests of the no hair theorem \cite{Cardoso:2016ryw, Carullo:2018sfu, Isi:2019aib, Cotesta:2022pci} and of the \ac{bh} area law \cite{Cabero:2017avf, Isi:2020tac}.

\Ac{nr} played a key role in this discovery, as it provides the only way to calculate the complete waveform for merging \acp{bh} from first principles \cite{Baumgarte:2010ndz}. Indeed, the need for large scale numerical computation to solve the two-body problem in general relativity was recognized already in the '70s with the pioneering work of Smarr et al.~\cite{Smarr:1976qy}. However, a concerted effort by a large community was necessary to achieve the \ac{bbh} breakthrough of the early 2000's \cite{Pretorius:2005gq, Campanelli:2005dd, Baker:2006ha, Baker:2006yw, Sperhake:2006cy, Bruegmann:2006ulg}. These early calculations revealed that the \ac{gw} signal from \ac{bbh} systems smoothly transitions from the inspiral portion, which is well described by post-Newtonian theory, to the ringdown, which is well described by \ac{bh} perturbation theory. The merger waveform itself was shown to be stunningly simple as it smoothly blends the inspiral waveform to the ringdown signal. Since then, \ac{nr} simulations have since significantly increased in sophistication and accuracy \cite{Scheel:2006gg, Szilagyi:2009qz, Thierfelder:2011yi, Loffler:2011ay, Babiuc:2010ze, Hilditch:2015aba, Bugner:2015gqa, Clough:2015sqa, Kidder:2016hev, Daszuta:2021ecf, Rashti:2023wfe, Daszuta:2023qdm, Rashti:2024yoc}, enabling the calibration of analytic waveform models, such as the \ac{eob} \cite{Buonanno:1998gg, Buonanno:2000ef, Ramos-Buades:2021adz, Pompili:2023tna, Ramos-Buades:2023ehm, Chiaramello:2020ehz, Akcay:2020qrj, Gamba:2021ydi, Nagar:2023zxh, Nagar:2024dzj, Nagar:2024oyk} and phenomenological \ac{imr} models~\cite{Ajith:2007qp, Ajith:2007kx, Ajith:2009bn, Santamaria:2010yb, Husa:2015iqa, Khan:2015jqa, Pratten:2020fqn, Estelles:2020osj, Estelles:2021gvs, Hamilton:2021pkf, London:2017bcn, Garcia-Quiros:2020qpx, Khan:2019kot, Hannam:2013oca, Schmidt:2014iyl, Khan:2018fmp, Pratten:2020ceb}, as well as purely data-driven models \cite{Blackman:2014maa, Blackman:2017dfb, Varma:2018mmi, Varma:2019csw, Williams:2019vub}. Large catalogs of waveforms have been produced and are continuously expanded as progressively larger regions of the binary parameter space are explored \cite{Hinder:2013oqa, Jani:2016wkt, Ferguson:2023vta, Mroue:2013xna, Boyle:2019kee, 2025arXiv250513378S, Healy:2017psd, Healy:2019jyf, Healy:2020vre, Healy:2022wdn, Huerta:2019oxn, Hamilton:2023qkv}.

Shortly after the announcement of GW150914, Lovelace and collaborators~\cite{Lovelace:2016uwp} presented \ac{nr} simulations \newtxt{targeted} to the event performed with two independent \ac{nr} codes: \texttt{LazEV} \cite{Campanelli:2005dd, Zlochower:2005bj} and \texttt{SpEC} \cite{Szilagyi:2009qz, Kidder:2016hev} representing state-of-the-art implementations of the two most successful approaches for the \ac{bbh} problem. The former implements the Baumgarte, Shapiro, Shibata, Nakamura (BSSN) formalism for the Einstein equations \cite{Nakamura:1987zz, Shibata:1995we, Baumgarte:1998te} with moving puncture gauge \cite{Brandt:1997tf, vanMeter:2006vi, Alcubierre:2002kk, Baker:2006ha, Campanelli:2005dd}, while the latter implements the generalized-harmonic formalism \cite{Lindblom:2005qh} with singularity excision. Ref.~\cite{Lovelace:2016uwp} demonstrated the excellent agreement between the theory predictions and between theory and experimental results, further establishing the interpretation of the signal GW150914 as being the result of a \ac{bbh} merger.

In this work, we revisit GW150914 using the new open source code \texttt{AthenaK} \cite{2024arXiv240916053S, Zhu:2024utz, Fields:2024pob}, showcasing the capabilities of new-generation \ac{nr} codes. While \texttt{AthenaK} is similar to \texttt{LazEV} in many ways, for example, in its use of the moving-puncture gauge, it also includes new developments. \texttt{AthenaK} uses the Z4c formulation of Einstein equations \cite{Bernuzzi:2009ex, Cao:2011fu}, which has been shown to result in better constraint preservation and accuracy compared to the original BSSN method. Most importantly, \texttt{AthenaK} is designed to run efficiently and at scale on large, GPU-accelerated machines, which are ${\sim}100$ times faster than leadership scale machines available in 2016. \texttt{AthenaK} development is fully public \cite{AthenaKweb} and we welcome contributions from the broader community. This paper is accompanied by a detailed step-by-step tutorial and a set of scripts needed to reproduce all simulations and results presented here \cite{AthenaK_BBH_Tutorial}, in the hope of further broadening participation in the \ac{nr} community.

The rest of this paper is organized as follows. We summarize the numerical techniques implemented in \texttt{AthenaK} and provide details on our initial data in Sec.~\ref{sec:methods}. Our results, as well as comparison with the data from Ref.~\cite{Lovelace:2016uwp} and a direct application to data analysis, are presented in Sec.~\ref{sec:results}. Finally, Sec.~\ref{sec:conclusions} is dedicated to discussion and conclusion. Unless otherwise stated, all quantities are given in natural units in which $M = G = 1$, where $M$ is the binary mass.

\section{Methods}
\label{sec:methods}
The computational kernels in \texttt{AthenaK} have been ported from \texttt{GR-Athena++} \cite{Daszuta:2021ecf} and are summarized in \cite{Zhu:2024utz}, where the differences with \texttt{GR-Athena++} are also discussed in detail. Here, we summarize the main feature of the code as they pertain the results presented here, but we refer to the previously mentioned works for more details.

We construct initial data using the pseudo-spectral elliptic solver \texttt{TwoPunctures} code~\cite{Ansorg:2004ds}. The parameters from the initial data are taken from Ref.~\cite{Lovelace:2016uwp} to match that used for the \texttt{LazEV} code. The Z4c equations are discretized in space using 6th order finite differencing, with advective derivatives along the shift vector $\beta^i$ lopsided by one grid point. The equations are then evolved in time with a low-storage 4th order Runge-Kutta method. Our computational grid covers a region with extent $[-2048, 2048]^3$. The base grid is covered with $128$ points in each direction.  To resolve the punctures, we employ 12 levels of adaptive mesh refinement (AMR; including the base grid)\acused{amr}, with the highest resolution region placed according to the position of the punctures. This results in a finest resolution of $\Delta = 0.0078125\, M$ corresponding to ${\sim}128$ points across the coordinate diameter of each \ac{bh}, after the initial relaxation of the gauge. Note that, because of the octree block-tree based nature of the \texttt{AthenaK} \ac{amr}, this results in a refinement structure that extends throughout the domain. However, we also enforce a minimum resolution of $\Delta = 0.25\, M$ within a (coordinate) radius $r = 200\, M$ to ensure that \acp{gw} are well resolved at the radii where we extract them $r \in \{25, 50, 100, 150, 200\}\, M$. The amplitude peak in the $(\ell,m)=(2,2)$ mode is reached at $t \simeq 2154\, M$, but we continue the simulations until time $2500\, M$. The total wall clock time was of ${\sim}130$ hours on 32 nodes of ALCF's Aurora (192 GPUs).

We extract \acp{gw} in two ways. First, we output the spin weighted multipolarly decomposed complex scalar $\Psi_4$ on spheres at different radii, e.g.,~\cite{Bishop:2016lgv}. This data is then integrated twice in time using the fixed-frequency integration method \cite{Reisswig:2010di} to obtain the strain data in the transverse-traceless gauge. This approach closely follow that of the \texttt{LazEV} team in \cite{Lovelace:2016uwp}. The other approach is through Cauchy-characteristic extraction \cite{Bishop:1998uk, Reisswig:2009us, Babiuc:2010ze, Handmer:2014qha, Moxon:2020gha, Moxon:2021gbv}. To this aim, we output metric data on a cylinder with $49\, M \leq r \leq 51\, M$, as outlined in Ref.~\cite{Rashti:2024yoc}. We then use the open-source code \texttt{SpECTRE} \cite{SpECTRECode} to compute waveforms at $\mathcal{I}^+$. We use the \texttt{scri} package \cite{scri_packages} to transform the resulting waveform to the super rest frame of the binary at a time shortly after the end of the junk radiation \cite{Boyle:2013nka, Boyle:2015nqa, Boyle:2014ioa, Mitman:2020pbt, MaganaZertuche:2021syq, Mitman:2022kwt}. Scripts to convert \texttt{AthenaK} data to the format needed by \texttt{SpECTRE} are included in Ref.~\cite{AthenaK_BBH_Tutorial}. \newtxt{In this work, we focus on the CCE data, which is free from finite-radius extraction artifacts and can also include GW memory effects. A detailed comparison between wave-extraction methods can be found in \cite{Reisswig:2009rx, Moxon:2020gha, Rashti:2024yoc}.}

\texttt{AthenaK} does not include an apparent horizon finder, however we record 3D metric data in cartesian boxes with diameter $4\, M$ centered at the location of the punctures. This data is then processed using \texttt{AHFinderDirect} \cite{Thornburg:2003sf} and \texttt{QuasiLocalMeasures} \cite{Dreyer:2002mx} from the \texttt{Einstein Toolkit} \cite{EinsteinToolkit:2024_11}. The minimal subset of the \texttt{Einstein Toolkit} used for our analysis was developed in the context of the \texttt{BlackHoles@Home} project \cite{Ruchlin:2017com}.

\section{Results}
\label{sec:results}

\subsection{Dynamics and remnant properties}

\begin{figure}
    \centering
    \includegraphics[width=0.5\linewidth]{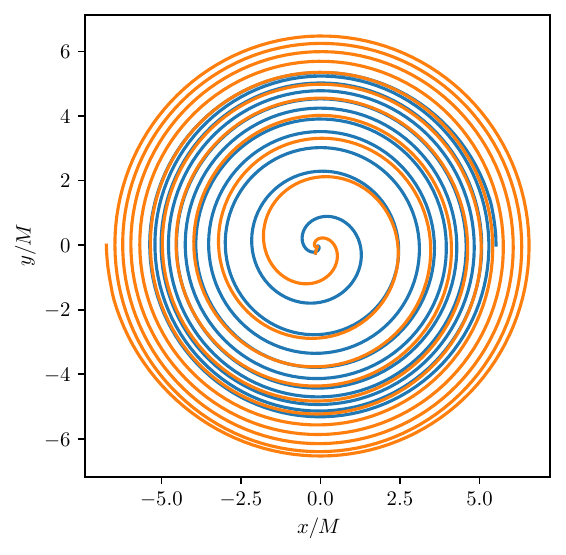}
    \caption{Puncture trajectories obtained by integrating $\dot{x}^i_A = -\beta^i(x_A)$. \newtxt{The primary BH's trajectory is in blue, while the secondary BH trajectory is in orange}. The eccentricity of the orbit, estimated from the trajectory, is $e= 8.3 \times 10^{-4}$.}
    \label{fig:trajectories}
\end{figure}

Our simulation starts from a coordinate separation of $11.8\, M$, corresponding to initial instantaneous orbital frequency of $M \omega = 0.20$, and evolve to merger in ${\sim}2050\, M$ over ${\sim}10$ orbits. Fig.~\ref{fig:trajectories} shows the trajectories of the two punctures from our simulation. We do not observe a significant drift of the center of mass of the system: \newtxt{by the time that the ringdown waveform has decayed to below the numerical noise ($t \simeq 2100\ M$) the coordinate position of the center of mass has moved only by $0.1\, M$ from the center (about 12 grid points).} The eccentricity, estimated from the puncture trajectories \newtxt{following the approach described in \cite{Tacik:2015tja, Tichy:2019ouu}}, is $e = 8.3 \times 10^{-4}$. \newtxt{This estimate is obtained by fitting a post-Newtonian expression for the orbital separation to the puncture's coordinate distance starting at $t = 100\, M$ and extending until the separation drops below $8\, M$. As such, $e$ should be taken as a measurement of eccentricity around $M \omega \sim 0.2$.} Note that the \acp{bh} spins are aligned to the orbital angular momentum, so no precession is present in the evolution.

\begin{figure}
    \centering
    \includegraphics[width=0.5\linewidth]{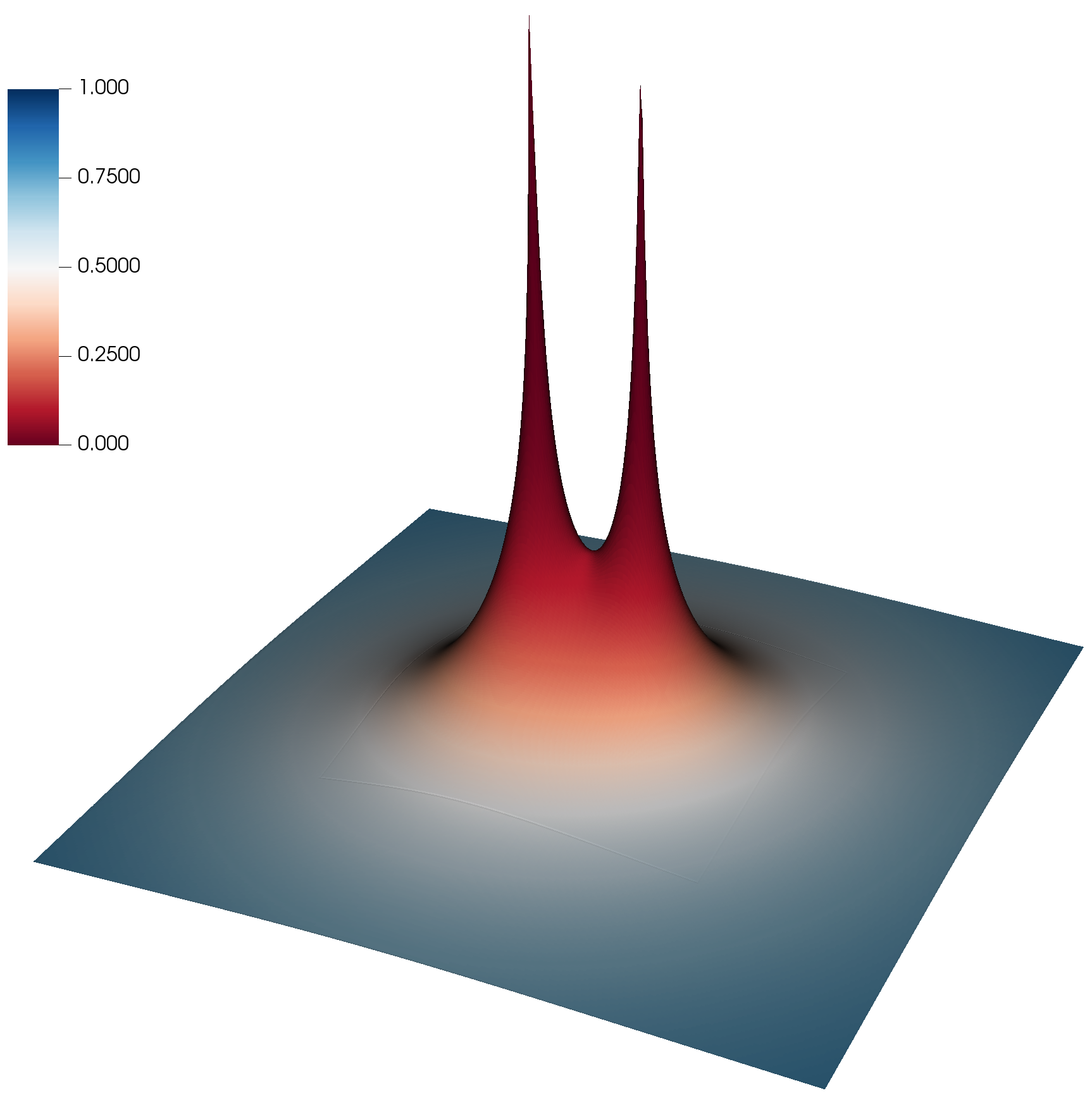}
    \caption{Conformal factor $\chi = \gamma^{-3}$, $\gamma$ being the determinant spatial metric, at the approximate time of merger $t = 2050\, M$.}
    \label{fig:eyecandy}
\end{figure}

Fig.~\ref{fig:eyecandy} shows the conformal factor on the orbital plane at the approximate time of merger. At this time a common apparent horizon has already been formed. The mass and dimensionless spin parameters of the remnant, computed using the isolated horizon formalism \cite{Ashtekar:2004cn}, are $M = 0.951948$ and $\chi = J/M^2 = 0.691914$. The relative difference with the highest resolution \texttt{SpEC} results reported in Ref.~\cite{Lovelace:2016uwp} are of $9 \times 10^{-5}$ and $2 \times 10^{-4}$, respectively. These imply that 4.81\% of the initial binary mass is radiated in \acp{gw}, \newtxt{consistent} with analytic and other numerical predictions~\cite{LIGOScientific:2016aoc, LIGOScientific:2016vlm, Lovelace:2016uwp}. The recoil velocity of the remnant \ac{bh}, estimated assuming momentum conservation and integrating the momentum flux in \acp{gw} with the \ac{cce} data, is of $138.68\, {\rm km}\ {\rm s}^{-1}$, in good agreement with Ref.~\cite{Lovelace:2016uwp}, as it is within $3\%$ of the highest resolution \texttt{SpEC} results. 

\subsection{Waveform and comparison against other catalogs}
\begin{figure}
    \centering
    \includegraphics[width=\linewidth]{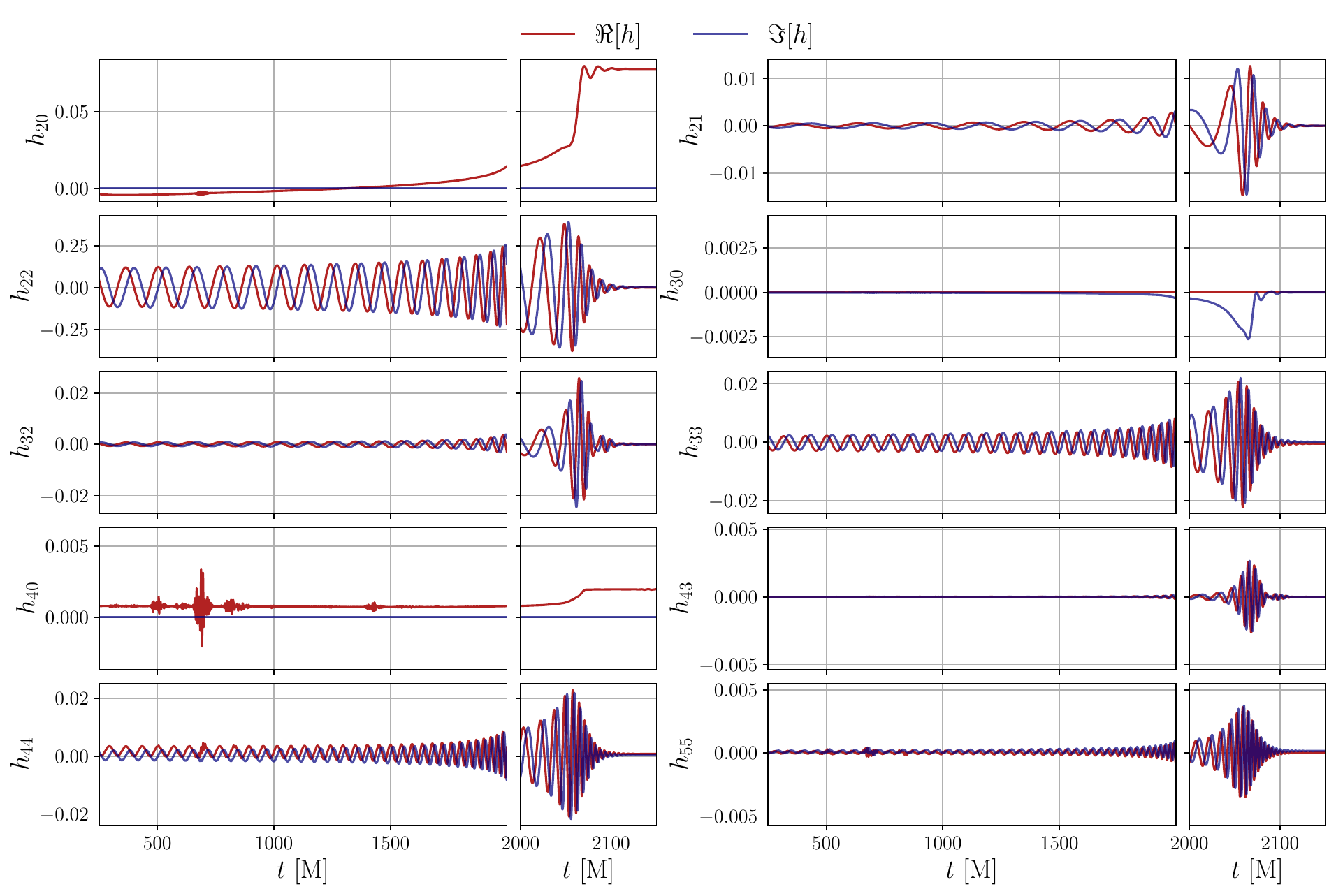}
    \caption{Real and imaginary parts of the waveform multipoles $h_{\ell m}$ extracted at $\mathcal{I}^+$ via \ac{cce}. Numerical noise is evident in the $(2,0),(4,0),(4,4)$ and $(5,5)$ modes around $t \simeq 700M$. Memory effects are also visible, particularly in the $m=0$ modes.}
    \label{fig:modes}
\end{figure}

A selection of waveform multipoles, extrapolated to $\mathcal{I}^+$ via \ac{cce}, is shown in Fig.~\ref{fig:modes}.
Our waveforms, notably, contain memory effects (see e.g. Ref.~\cite{Mitman:2024uss} for a review), which are especially apparent in the $m=0$ modes shown. Such effects would cause a displacement in the arms of a detector 
after a \ac{gw} has passed through. In our modes, they appear as a ``global offset'' that slowly grows
during the inspiral, and does not taper to zero beyond merger.
One additional feature related to memory is visible in the imaginary part of the $\ell=3, m=0$ mode: this is the so-called ``spin memory'' effect, related to the change in relative separation of observers with initial relative velocity~\cite{Nichols:2017rqr, Mitman:2020pbt}.
It is possible to see noise events in our \ac{gw} strain data, a prominent example is around $t = 750\, M$ in the $h_{40}$ mode. This noise is possible due to back scattering of waves at \ac{amr} boundaries towards the extraction radius. \newtxt{A comparison with data from a higher resolution \texttt{AthenaK} simulation is discussed in \ref{sec:appendix}.}

\begin{figure}
    \centering
    \includegraphics[width=1.\linewidth]{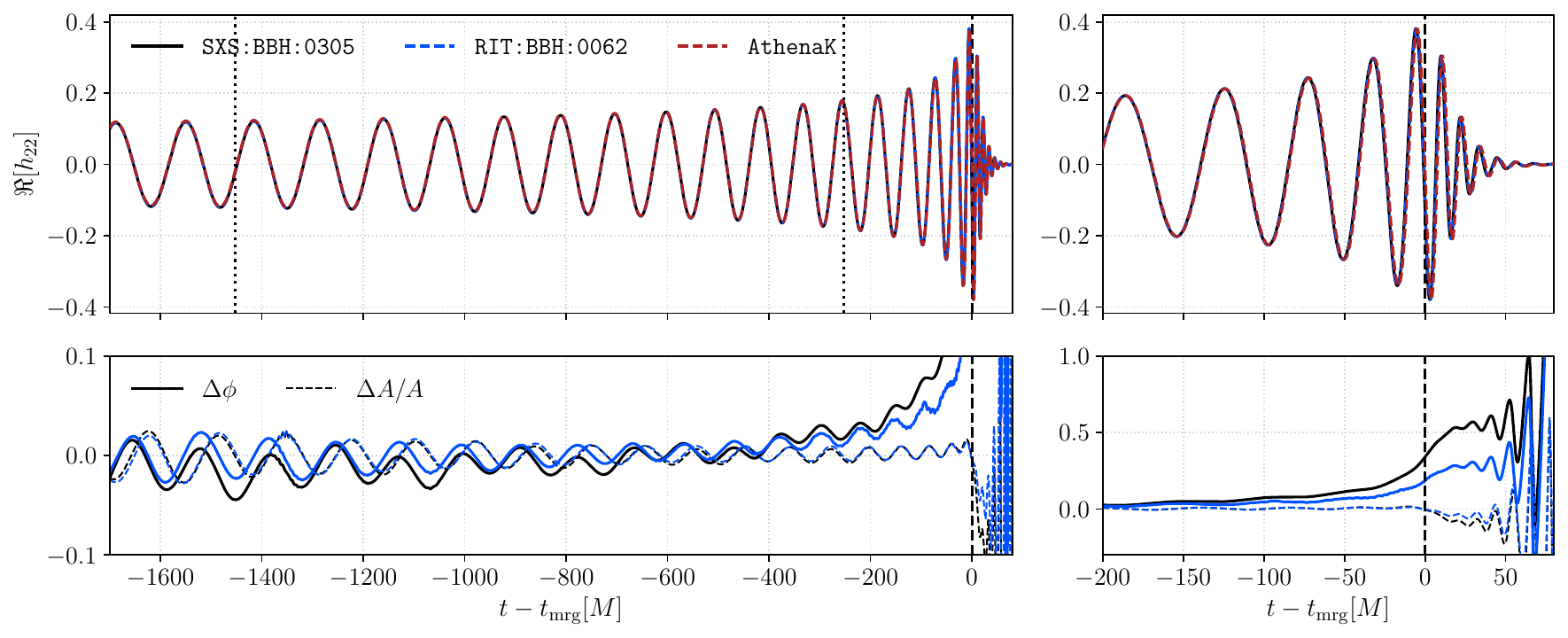}
    \caption{Comparison of the dominant $\ell=m=2$ waveform multipole obtained from our targeted simulation (red dashed) against the same quantity from the $\texttt{SXS:BBH:0305}$ (black) and $\texttt{RIT:BBH:0062}$ (blue dashed) simulations. The phase and amplitude differences are reported in the bottom panel, and remain small throughout the entire evolution, with a maximum $\Delta\phi \sim 0.35$ and $\Delta A/A \sim 0.4\%$ around merger.}
    \label{fig:comparison}
\end{figure}

Figure~\ref{fig:comparison} presents a time domain phasing comparison between the dominant quadrupolar mode obtained in this work and the same quantity extracted from the \texttt{SXS:BBH:305} and \texttt{RIT:BBH:0062} simulations, which are respectively the highest resolution calculations obtained with \texttt{SpEC} and \texttt{LazEV} of Ref.~\cite{Lovelace:2016uwp}.
Following e.g. Ref.~\cite{Rashti:2024yoc} we align the modes between the times $t_i, t_f$ demarcated by vertical dotted lines. This is done by determining the time and phase shifts ($\Delta t$, $\Delta \phi$) such that the following quantity
\begin{equation}
    \chi^2 = \int_{t_i}^{t_f} [\phi_{\rm X}(t+\Delta t) - \phi_{\rm AthenaK} - \Delta \phi]^2 dt
\end{equation}
is minimized. In the above, $\phi_{X}$ denotes the GW phase of the $(2,2)$ mode 
obtained with the code $\rm X = \{\texttt{SpEC}, \texttt{LazEv}\}$.
Qualitatively, the visual agreement among the waveforms is excellent throughout the entire coalescence. Quantitatively, the cumulative phase difference at merger $\Delta\phi$ amounts to $\approx 0.35$ and $0.2$ rad between the \texttt{SXS}-\texttt{AthenaK} and \texttt{RIT}-\texttt{AthenaK} waveforms, respectively. 
The amplitude relative difference, $\Delta A/A$, remains almost constant until merger, at a value $0.2\%- 0.4\%$ depending on the simulation considered. These values are comparable to those obtained by current state of the art models, calibrated to \ac{nr} (see e.g. Fig.~1 of Ref.~\cite{Gamba:2025qfg}).

\subsection{Application to GW150914}
The waveform multipoles $h_{\ell m}$ obtained via our targeted simulations can be used directly to estimate some of the source properties of GW150914. \newtxt{In particular, we can use the numerical relativity data to estimate source parameters that are extrinsic to the binary (sky position, orientation), because it is possible to sample over these parameters without generating new simulation data.} Indeed, while mass ratio and (dimensionless) spins are set at the beginning of the simulation, the total mass $M$, inclination $\iota$, reference phase $\phi_{\rm ref}$, polarization $\psi$, luminosity distance $d_L$ , right ascension and declination $\alpha$, $\delta$ can be freely
varied to project the plus and cross polarizations $h_{+}, h_{\times}$ on the detectors:
\begin{equation}
    h_+ - i h_{\times} = \frac{M}{d_L}\sum_{\ell m} h_{\ell m} {}_{-2}Y_{\ell m}(\iota, \phi_{\rm ref})
\end{equation}
\begin{equation}
    h = h_+ F_+(\psi, \alpha, \delta) + h_{\times} F_{\times}(\psi, \alpha, \delta) \, ,
\end{equation}
where $F_+$ and $F_\times$ are the antenna pattern functions.

\begin{figure}
  \centering
  \includegraphics[width=\textwidth]{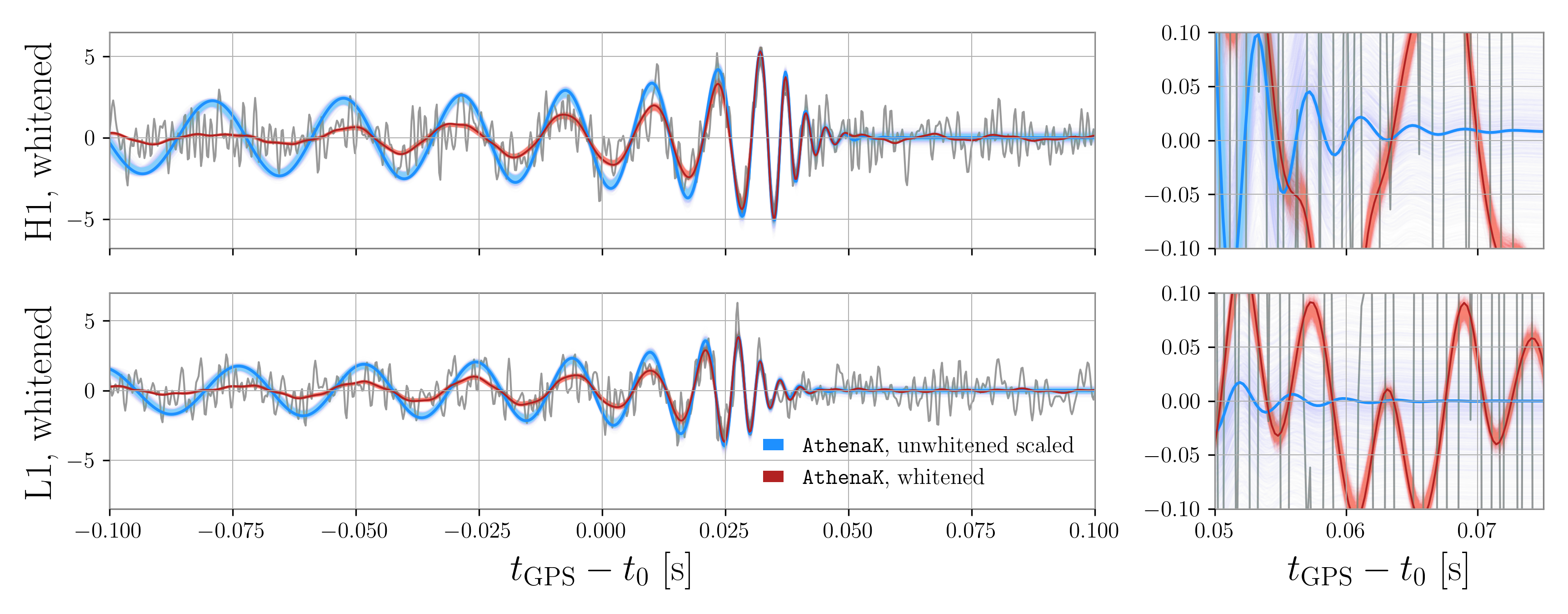}
  \caption{Reconstructed raw (blue) and whitened (red) waveforms, obtained performing PE with the modes $h_{\ell m}$ from the \texttt{AthenaK} simulation, overlaid to the \newtxt{whitened} strain data from the Hanford (top) and Livingston (bottom) detectors around the time of GW150914, $t_0$. The dark red waveform corresponds to the maximum likelihood parameters; error bands around it are instead obtained considering $2000$ waveforms from the analysis posterior, and are representative of the localization/mass uncertainty. The unwhitened waveform data clearly shows the impact of memory effects (right panels), with the polarizations not tapering to zero in H1.}
  \label{fig:reconstructed}
\end{figure}

\begin{figure}
  \centering
  \includegraphics[width=0.55\linewidth]{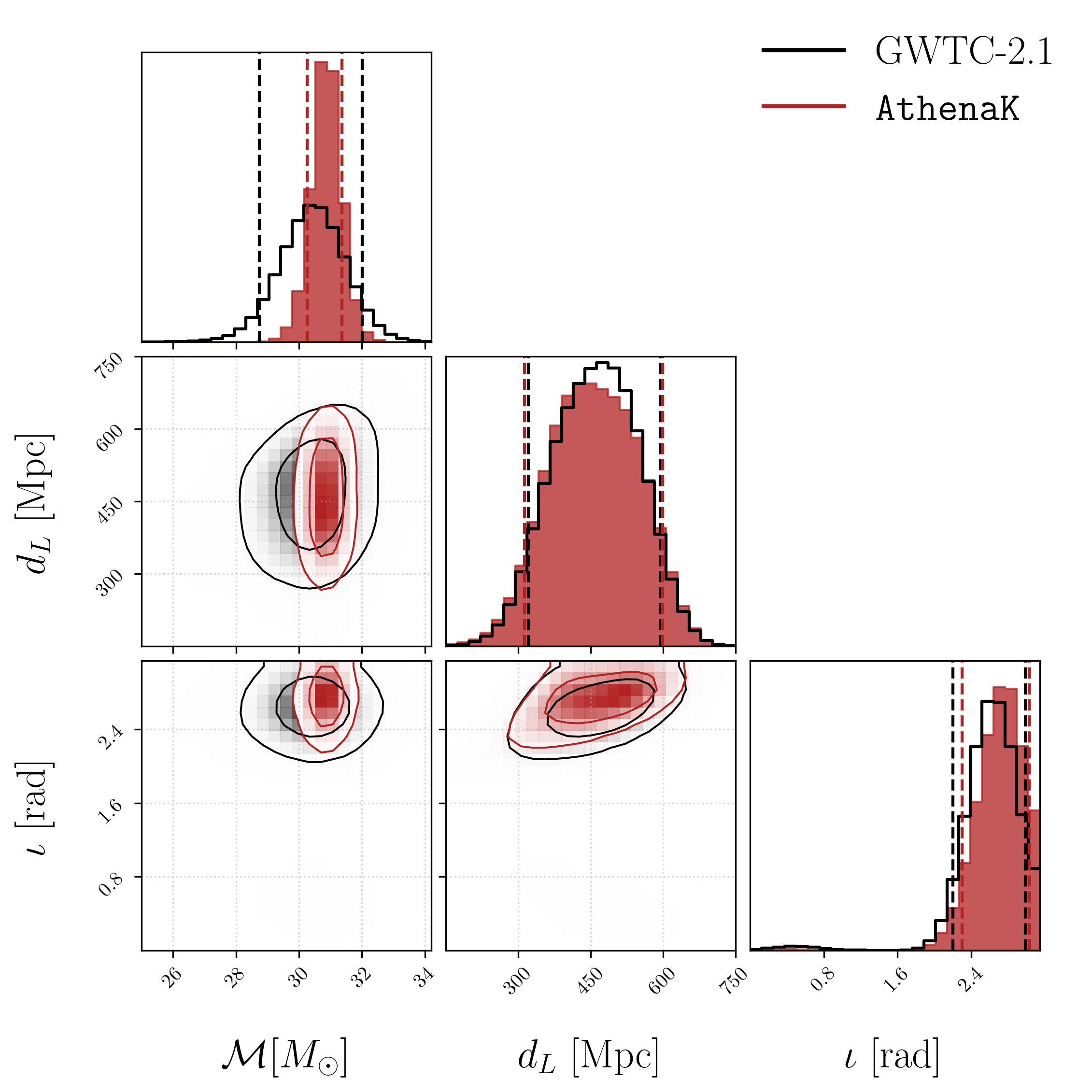}
  \caption{Posterior distribution of chirp mass $\mathcal{M}$, luminosity distance $d_L$ and inclination $\iota$ recovered by analyzing the data of GW150914 using the waveform produced by our \texttt{AthenaK} simulation (red) compared to the full analysis of GWTC-2.1 performed with the \texttt{IMRPhenomXPHM} model~\cite{Pratten:2020ceb, LIGOScientific:2021usb} (black). The estimates obtained from our simulation agree with those obtained from the semi-analytical model within $90\%$ credibility.}
  \label{fig:gw150914_corner}
\end{figure}

We then re-analyze GW150914 using the \texttt{bilby} parameter estimation library~\cite{Ashton:2018jfp}. We consider $(2, 0), (2, \pm 2), (2,\pm 1), (3,\pm 3), (3, \pm 2), (4,\pm 4), (4,\pm 3), (4,\pm 2), (5,\pm 5)$ waveform multipoles and sample over the parameters listed above using standard priors used in \ac{gw} analysis, listed in Tab.~\ref{tab:priors}.
We download data from the Gravitational Wave Open Science Center (GWOSC)~\cite{LIGOScientific:2019lzm, KAGRA:2023pio}, consider $8$ seconds around the time GPS of event $t_0 = 1126259462.391$~s and the frequency range $f \in [35, 896]$ Hz. 
Figure~\ref{fig:reconstructed} shows the visual agreement of the reconstructed waveform and its uncertainty with the whitened detector data: our targeted simulation clearly captures the strain induced by the incoming \ac{gw} signal. Figure~\ref{fig:gw150914_corner}, instead, displays the posterior distributions of chirp mass $\mathcal{M}$, luminosity distance $d_L$ and inclination $\iota$ we obtain. We find $\mathcal{M} = 30.7^{+0.6}_{-0.5} M_{\odot}$, $d_L = 460^{+140}_{-140}$ Mpc, $\iota = 2.7^{+0.3}_{-0.4}$ rad. These results are consistent with estimates obtained using state-of-the-art semi-analytic waveform models~\cite{Pratten:2020ceb, Ossokine:2020kjp}, albeit more constrained. This difference is readily explained by the fact that, in our analysis, the mass ratio and component spins of the binary are fixed, thereby reducing the impact of parameter correlations.

\begin{table}[]
    \centering
    \begin{tabular}{l l l}
        \br
        Parameter name   & Prior & Range \\
      \mr
      Total mass $M [M_{\odot}]$   & Uniform &$[40, 100]$ \\
      Luminosity distance $d_{\rm L}$  [Mpc]  & Uniform in source frame &$[50, 2000]$ \\
      Inclination $\iota$ [rad] & Sine & $[0, \pi]$ \\
      Polarization $\psi$ [rad] & Uniform (periodic) & $[0, \pi]$ \\
      Reference phase $\phi_{\rm ref}$ [rad] & Uniform (periodic) & $[0, 2\pi]$ \\
      Right ascension $\alpha$ [rad] & Uniform (periodic) & $[0, 2\pi]$ \\
      Declination $\delta$ [rad] & Cosine & $[-\pi/2, \pi/2]$\\
      Coalescence time $t_{\rm c}$ [s] & Uniform & [$t_0 - 6$, $t_0 + 2$] \\
    \br
    \end{tabular}
    \caption{Priors on intrinsic ($M$, $\iota$) and extrisinc ($d_L, \psi, \phi_{\rm ref}, \alpha, \delta, t_{\rm c}$) parameters employed for the re-analysis of GW150914 with the $\texttt{AthenaK}$ waveform modes.}
    \label{tab:priors}
\end{table}

\section{Conclusions}
\label{sec:conclusions}
We have revisited the \ac{bbh} merger in GW150914 using the new, GPU-accelerated, \ac{nr} code \texttt{AthenaK} \cite{2024arXiv240916053S, Zhu:2024utz, Fields:2024pob}. This work serves as first cross-validation of \texttt{AthenaK} with established \ac{nr} codes  \texttt{LazEV \cite{Campanelli:2005dd, Zlochower:2005bj}} and \texttt{SpEC} \cite{Szilagyi:2009qz, Kidder:2016hev}, as well as with real data. It also serves as demonstration of our pipeline for \ac{gw} astronomy, which is made fully available to the community \cite{AthenaK_BBH_Tutorial}. 

We have computed remnant properties using the isolated horizon formalism \cite{Ashtekar:2004cn} and found excellent agreement with the \texttt{LazEV} and \texttt{SpEC} results presented in Ref.~\cite{Lovelace:2016uwp}. We have demonstrated the use of \texttt{AthenaK} world tube data in combination with \texttt{SpECTRE} \cite{SpECTRECode} to compute \acp{gw} at $\mathcal{I}^+$  including memory effects, which were neglected in previous simulations of GW150914. We have found agreement between the dominant $(\ell,m)=(2,2)$ mode of the \ac{gw} signal obtained with this approach with previous results by Ref.~\cite{Lovelace:2016uwp}, with flat residuals in amplitude and phase during most of the inspiral. The cumulative dephasing between \texttt{AthenaK} and \texttt{LazEV}/\texttt{SpEC} is of $\Delta \phi \simeq 0.2-0.35$ radians at merger, while the relative amplitude difference is of $\Delta A/A \simeq 0.2\% - 0.4\%$.
Finally, we have applied our simulation to the analysis of real GW data, performing a parameter estimation study on GW150914.
We found very good agreement with the previous results from the LIGO-Virgo-Kagra collaboration, estimating $\mathcal{M} = 30.7^{+0.6}_{-0.5} M_{\odot}$, $d_L = 460^{+140}_{-140}$ Mpc and $\iota = 2.7^{+0.3}_{-0.4}$ rad.

The main limitation of this work is that we have only considered a single grid resolution with \texttt{AthenaK}. Future work is also needed to address possible gauge effects in the waveform comparison between \texttt{AthenaK} and other codes. Work along these lines is already ongoing and will be reported in the future. 

\section*{Acknowledgements}
We thank Zachariah Etienne for having provided us the minimal version of the
\texttt{Einstein Toolkit} used for the black hole horizon analysis, and the
\texttt{SpECTRE} developers for their assistance with \texttt{SpECTRE}'s
Cauchy-characteristic extraction module. DR gratefully acknowledge the
hospitality and support from Sorbonne Universit\'e,  where part of this work was
carried out. This work was supported by NASA through Awards 80NSSC21K1720 and
80NSSC25K7213, and by the National Science Foundation under Grant
No.~PHY-2020275 (Network for Neutrinos, Nuclear Astrophysics, and Symmetries
(N3AS)). This research used resources of the Argonne Leadership Computing
Facility, which is a DOE Office of Science User Facility supported under
Contract DE-AC02-06CH11357. Computer time was provided by the INCITE program.
Any opinions, findings, and conclusions or recommendations expressed in this
material are those of the authors and do not necessarily reflect the views of
the Department of Energy, NASA, or the National Science Foundation.

\section*{References}
\bibliographystyle{iopart-num}
\providecommand{\newblock}{}

\appendix

\newpage
\section{Higher resolution simulation}
\label{sec:appendix}
\newtxt{In this section we validate the \texttt{AthenaK} data using results from a higher-resolution simulation, with otherwise identical setup. In particular, we perform an additional simulation with resolution of $192$ points in the coarsest refinement level, as opposed to $128$ of the baseline simulation. This results in a finest resolution of $\Delta_{192} = 0.0052083$.}

\begin{figure}
    \centering
    \includegraphics[width=\linewidth]{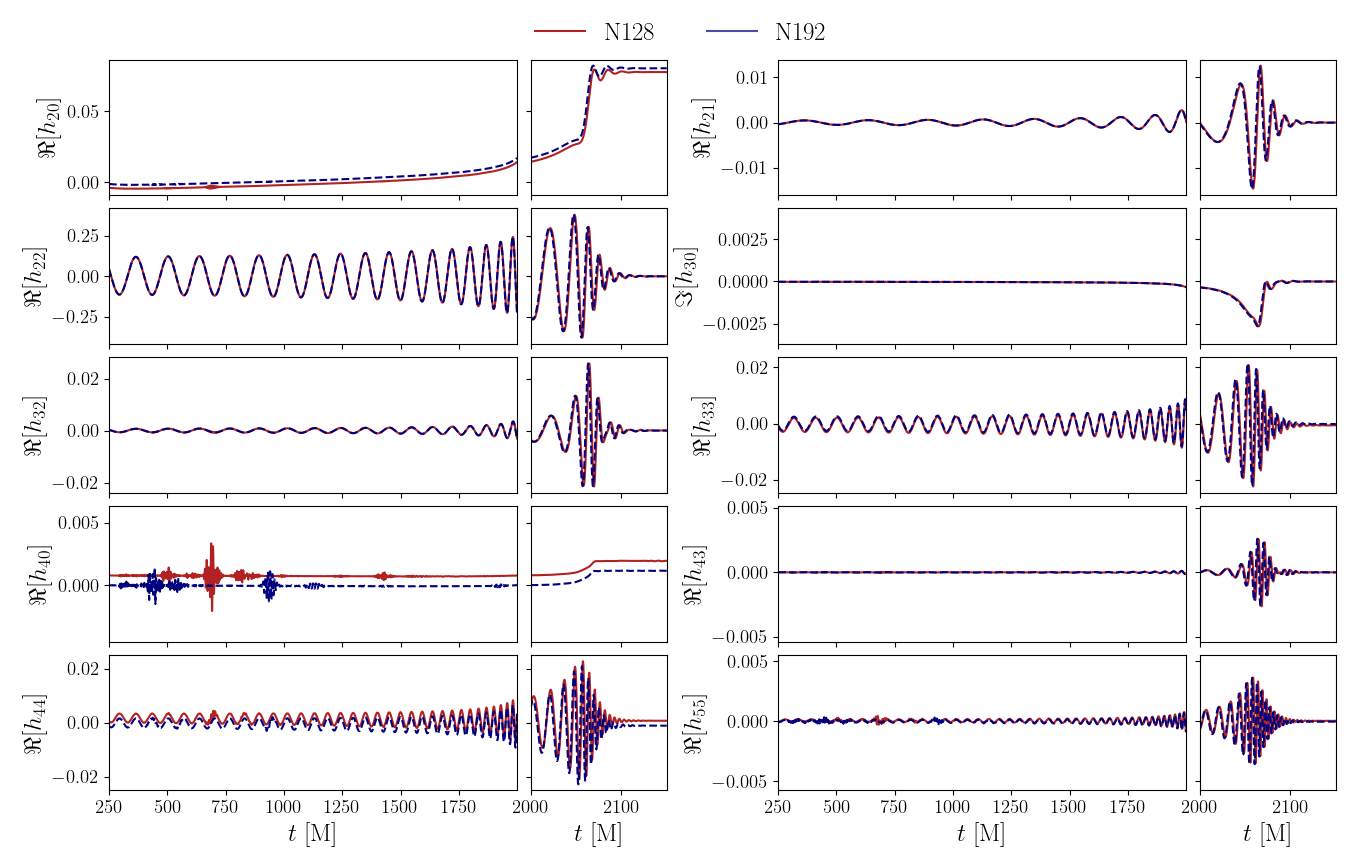}
    \caption{\newtxt{Real (or imaginary) parts of the waveform multipoles $h_{\ell m}$ extracted at $\mathcal{I}^+$ via \ac{cce} for the $N=192$ simulation, compared to the lower resolution one with $N=128$.}}
    \label{fig:modes_res}
\end{figure}

\newtxt{
The modes obtained from the higher-resolution simulation are presented in
Fig.~\ref{fig:modes_res}. Notably, differences among the two resolutions also
impact the determination of the superrest frame of each binary. As such, some of
the modes dominated by memory contributions, e.g. the $(\ell = 4, m=0)$ mode,
appear rather different as the resolution is increased. However, we obtain
overall good agreement between the two resolutions in the phase and amplitude
evolution of most of the modes. For example, the phase difference for the $(\ell
= 2, m=2)$ mode at merger is $\Delta\phi_{22} = 0.38$ and the relative amplitude
difference is $\Delta A_{22} / A_{22} = 0.2\%$. Compared to the SXS (RIT)
simulation shown in Fig.~\ref{fig:comparison}, this resolution achieves a phase
difference of $\Delta\phi = 0.12~(0.17)$ rad, and a relative amplitude
difference of $\Delta A_{22} / A_{22} = 0.01\%$ ($0.02\%$).
}

\end{document}